\documentstyle[11pt]{article}
\begin{document}

\title{Probing flavor changing interactions in photon-photon collisions
\footnote{This work was supported by National Natural Science Foundation of
          China}}

\author{{
  Jiang Yi$^{b}$, Zhou Mian-Lai$^{b}$, Ma Wen-Gan$^{a,b,c}$,
  Han Liang$^{b}$, Zhou Hong$^{b}$ and Han Meng$^{b}$}\\
{\small $^{a}$CCAST (World Laboratory), P.O.Box 8730, Beijing 100080,
China.}\\
{\small $^{b}$Department of Modern Physics, University of Science
and Technology}\\
{\small of China (USTC), Hefei, Anhui 230027,
             China. } \\
{\small $^{c}$Institute of Theoretical Physics, Academia Sinica,}\\
{\small P.O.Box 2735, Beijing 100080, China.} }
\date{}
\maketitle

\begin{center}\begin{minipage}{100mm}

\vskip 5mm
\begin{center} {\bf Abstract}\end{center}
\baselineskip 0.3in
{We examine the subprocess $\gamma\gamma \rightarrow t\bar{c}+\bar{t}c$
at electron-positron colliders in the two-Higgs-doublet model with
flavor-changing scalar couplings, where all the one-loop contribuions
are considered, and the results are applicable to the whole mass
range of the weakly coupled Higgs bosons. Because of the heavy top quark mass,
this process is important in probing the flavor-changing
top-charm-scalar vertex and could be detectable at the Next
Linear Collider, if the values of the parameters are favorable.
The results show that this process is more promising than the
direct $e^{+} e^{-}$ process for discovering flavor changing
scalar interactions. } \\

\vskip 5mm
{PACS number(s):13.65.+i, ~11.30.Hv, ~12.60.Fr, ~14.65.Ha}
\end{minipage}
\end{center}

\newpage
\noindent
{\Large{\bf I. Introduction}}

\begin{large}
\baselineskip 0.35in

The experimental data have shown that the flavor changing scalar interactions
(FCSI) involving the light quarks are strongly suppressed. This leads to
the suppression of the flavor changing neutral currents(FCNC) which is an
important feature of the standard model(SM) which was explained in terms of
the GIM mechanism. In the commonly used two-Higgs-doublet models(THDM), the
absence of the FCSI at the tree level can be assured, if the natural flavor
conservation(NFC) condition\cite{NFC} is valid, by imposing
discrete symmetries on the model. Then the THDM model with NFC condition
can be subdivided into two modes, i.e., Model I and Model II. In Model I,
both the up and down type quarks get their masses from the same Higgs
doublet, and in Model II the quarks get their masses from different doublets.
Two years ago the CDF and D0 collaborations found that the top quark has a very
large mass (world average value: $m_{t} = 175.6 \pm 5.5~GeV$)\cite{CDF}.
This extraordinary mass scale of the top quark has many important
implications pertaining to many outstanding issues in theoretical particle physics.
One of these consequences is that FCSI at the tree level would exist at high
mass scales. The measurement of FCSI involving top quark would provide an
important test for the discrimination of various models. As Cheng, Sher
and other authors\cite{fs}\cite{bfs}\cite{chengs}\cite{M3} pointed
out, since the Yukawa couplings are typically related to the masses
of the fermions participating at the vertices, it is rather natural to
expect having Yukawa couplings for the FCSI instead of placing the constraints
due to NFC on the theory. If the Yukawa couplings are proportional to the
quark masses, low energy limits on FCNC's may be evaded because the flavor
changing couplings to the light quarks are small, and the suppression of
the FCSI involving light quarks can be automatically satisfied.
Then the imposition of discrete symmetries to insure that the NFC condition
is satisfied, which is normally
invoked in commonly used two-Higgs-doublet models to prevent the FCSI at
the tree level, is therefore unnecessary. We call such
THDM as the third mode of THDM(i.e., THDM III) \cite{M3}. In the framework
of this model the effects of FCSI involving the heavy quark will be enhanced.

If the philosophy of the THDM III is correct, one would expect that large
effects of the FCSI could manifest themselves in the cases involving the
massive top quark. Therefore testing the existence of the flavor changing
scalar interactions involving top quarks is a promising task for future
colliders. Recently, D. Atwood et al. \cite{atwood} presented results of a
calculation
for the process $e^{+}e^{-}\rightarrow t \bar{c}$(or $\bar{t}c$) in the THDM
III, and they got $R^{tc}/\lambda^{4}$ to be in the order of $10^{-5}$
with proper parameters. They stressed that this experimental signal
is very clean and that FCSI can lead to measurable effects.
As we know, the future Next Linear Collider(NLC)
is designed as a $500~GeV$ $e^{+}e^{-}$ collider with an integral
luminosity of the order of $10~fb^{-1}$ per year. There is also a possibility
that the NLC may be operated in $\gamma \gamma$ collision mode, then it
provides another facility in top physics research with a cleaner environment.
The subprocess $\gamma\gamma \rightarrow h^{0}, A^{0} \rightarrow
t\bar{c}+\bar{t}c$ was first studied by Hou and Lin\cite{houlin}. They
pointed out that at the NLC we can also use this processs to study FCSI.
There they presented results of calculations for the specific case where
on-mass-shell neutral Higgs bosons with the masses in the range
$ 200~GeV < m_{h^{0},A^{0}} < 2~m_{t} \simeq 350~GeV$
are produced from $\gamma \gamma$ fusion and the photon
helicities have the average value $<\lambda \lambda^{'}>\sim +1$.
They predicted that one can get $10^{2}~\sim~10^{3}$ raw
events per year with $50~fb^{-1}$ luminosities in NLC operated
in $\gamma \gamma$ collision mode.

In this paper we present the complete one-loop calculation for the subprocess
$\gamma \gamma \rightarrow t\bar{c}$(or $\bar{t}c$) at the
$O(m_{t} m_{c}/m_{W}^{2})$ order in the THDM III, and the results are
applicable to the whole mass range for weakly coupled Higgs bosons.
The production rates of $e^{+}e^{-} \rightarrow \gamma \gamma \rightarrow
t\bar{c}+\bar{t}c$ are also given for the NLC energy range.
It shows that this process is more promising than the straight
$e^{+}e^{-}$ process for probing FCSI.
The paper is organized as follows: The details of calculation are given
in Sec. II. In Sec. III there are numerical results and discussion and
a short summary. Finally, the explicit expressions used
in the paper are collected in appendix.

\vskip 5mm
\noindent
{\Large{\bf II. Calculation}}
\vskip 5mm

In the third type of the two-Higgs-doublet model, the up-type and down-type
quarks are allowed simultaneously to couple to more than one scalar doublet.
We consider the THDM has two scalar $SU(2)_{w}$ doublets, $\phi_{1}$
and $\phi_{2}$:

$$
\phi_{1}=
\left(
\begin{array}{l}
\phi_{1}^{+} \\
\phi_{1}^{0}
\end{array}
\right) ,
~~\phi_{2}=
\left(
\begin{array}{l}
\phi_{2}^{+} \\
\phi_{2}^{0}
\end{array}
\right),
\eqno{(1)}
$$
with Lagrangian
$$
{\cal L}_{\phi}=D^{\mu}\phi^{+}_{1}D_{\mu}\phi_{1}+
               D^{\mu}\phi^{+}_{2}D_{\mu}\phi_{2}-
               V(\phi_{1},\phi_{2}),
\eqno{(2)}
$$
where $V(\phi_{1},\phi_{2})$ is the general potential which is consistent
with the gauge symmetries. Since there is no global symmetry that
distinguishes the two doublets in THDM III, we can assume that
$$
<\phi_{1}>=
\left(
\begin{array}{l}
0 \\
v/\sqrt{2}
\end{array}
\right) ,
~~<\phi_{2}>=0
\eqno{(3)}
$$
where $v \simeq 246~GeV$. The physical spectrum of Higgs bosons consists of two
scalar neutral bosons $h^{0}$ and $H^{0}$, one pseudoscalar neutral boson
$A^{0}$ and two charged Higgs $H^{\pm}$,
$$
\begin{array}{l}
H^{0}=\sqrt{2}[(Re\phi_{1}^{0}-v)\cos{\alpha}+Re\phi_{2}^{0}\sin\alpha],\\
h^{0}=\sqrt{2}[-(Re\phi_{1}^{0}-v)\sin{\alpha}+Re\phi_{2}^{0}\cos\alpha],\\
A^{0}=\sqrt{2}(-Im\phi_{2}^{0}).
\end{array}
\eqno{(4)}
$$
The masses of the five neutral and charged Higgs bosons and the mixing angle
$\alpha$ are free parameters of the model. The Yukawa couplings to quarks
are\cite{Luke},
$$
{\cal L}^{Q}_{Y}=\lambda^{U}_{ij}\bar{Q_{i}}\tilde{\phi_{1}}U_{j}+
                \lambda^{D}_{ij}\bar{Q_{i}}\phi_{1}D_{j}+
                \xi^{U}_{ij}\bar{Q_{i}}\tilde{\phi_{2}}U_{j}+
                \xi^{D}_{ij}\bar{Q_{i}}\phi_{2}D_{j},
\eqno{(5)}
$$
where the first two terms give masses of the quark mass
eigenstates, and $\xi^{U}_{ij}$ and $\xi^{D}_{ij}$ are the $3\times 3$
matrices which give the strength of the flavor changing neutral scalar
vertices. The $\xi$s are all free parameters and can be constrained by the
experimental data. If we neglect CP violation, the $\xi$s are all
real. In this paper we use the Cheng-Sher Ansatz(CSA)\cite{chengs} and let
$$
\xi_{ij}\sim \frac{\sqrt{m_{i}m_{j}}}{v}.
$$
And we can parametrize the Yukawa couplings as
$$
\xi_{ij}=g \frac{\sqrt{m_{i}m_{j}}}{m_{W}} \lambda.
\eqno{(6)}
$$
Comparing it with the usual gauge couplings of $SU(2)\times U(1)$,
one has $\lambda=\frac{1}{\sqrt{2}}$. In our calculation we use
$\lambda=\frac{1}{\sqrt{2}}$ and note that there is no stringent bound on
the coupling factor $\lambda$ theoretically.

This process can be produced via one-loop diagrams, and the Feynman diagrams
are given in figure 1(a) and figure 1(b), where the contribution of
neutral Higgs and charged Higgs one-loop diagrams are given
respectively. The diagrams exchanging the two external $\gamma\gamma$
lines are not shown in Fig.1(a) and Fig.1(b). Fig.1(a)(1 $\sim$ 12) and
Fig. 1(b)(1 $\sim$ 6) are the
self-energy diagrams, Fig. 1(a)(13 $\sim$ 20) and Fig. 1(b)(7 $\sim$ 10,
13 $\sim$ 16) are the vertex correction diagrams, Fig. 1(a)(25 $\sim$ 28) are
the s-channel diagrams, Fig.1(b)(19) is the quartic vertex diagram and
Fig. 1(a)(20 $\sim$ 24) and Fig. 1(b)(11 $\sim$ 12, 17 $\sim$ 18, 20 $\sim$ 21)
are the box diagrams. There is no tree-level contribution,
therefore the proper vertex counterterm cancels the counterterms of the
diagrams with external legs. That is to say the evaluation can be simply carried
out by summing all unrenormalized reducible and irreducible diagrams.

To simplify the calculation we set $\alpha=0$ as adopted in
Refs.\cite{atwood} and \cite{Luke} and let all scalar bosons be degenerate,
i.e., $m_{h^{0}}=m_{A^{0}}=m_{H^{\pm}}=M_{s}$ where $M_{s}$ is the common
scalar mass. The contribution from the coupling involving $H^{0}$ is suppressed
due to $\alpha=0$.

In the calculation for the s-channel diagrams(Fig.1.(a)(25 $\sim$ 26)), we
take into account the width effects of the
$h^{0}$ and $A^{0}$ propagators. As we know, the decays of $h^{0}$ to
WW and ZZ are suppressed, because
of the factor $\sin\alpha$ in the $h^{0}WW$, and $h^{0}ZZ$ couplings
and $h^{0}$ decay to $A^{0}A^{0}$ is also forbidden due to the case of
the degenerate masses of $h^{0}$ and $A^{0}$. Note that the pesudoscalar
$A^{0}$ does not couple with gauge boson pair. Therefore only the
decays of $h^{0}$ and $A^{0}$ to final states $q_{i}\bar{q_{j}}$ need
to be considered, where $q_{i}$ and $q_{j}$ represent quarks of
flavor i and j respectively. The decay width for the scalar $h^{0}$ can
be written as\cite{width}
$$
\Gamma (h^{0}\rightarrow q\bar{q})=\frac{3 g^{2} m_{h^{0}}}{32\pi M_{W}^{2}}
m_{q}\left(1-4m_{q}^{2}/m_{h^{0}}^{2}\right)^{3/2}
$$
and
$$
\Gamma (h^{0}\rightarrow t\bar{c}+\bar{t}c)=\frac{3g^{2}m_{h^{0}}}
{32\pi M_{W}^{2}}\cdot 2m_{t}m_{c}\left[1-(m_{t}+m_{c})^{2}/m_{h^{0}}^{2}
\right]^{3/2}\times \left[1-(m_{t}-m_{c})^{2}/m_{h^{0}}^{2}\right]^{1/2}.
\eqno{(7)}
$$
The decay width for the pesudoscalar $A^{0}$ boson can be represented
by exchanging exponents $3/2 \leftrightarrow 1/2$
and $m_{h^{0}} \leftrightarrow m_{A^{0}}$ in Eq.(7). When
$m_{t} + m_{c} < M_{s} < 2 m_{t}$, the dominant decay modes of $h^{0}$ and
$A^{0}$ are $h^{0}, A^{0} \rightarrow c\bar{c}, b\bar{b},t\bar{c}+\bar{t}c$,
whereas when $M_{s} > 2m_{t}$, the final state $t\bar{t}$ decay channel
is open, and their decay widths are rather large due to the large
masses of $M_{s}$ and $m_{t}$.

We denote $\theta$ as the scattering angle between one of the photons and
the final top quark. Then in the center-of-mass(CMS) we express all the
four-momenta of the initial and final particles by means of the total
energy $\sqrt{\hat{s}}$ and the scattering angle $\theta$. The four-momenta
of top quark and charm quark are $p_{1}$ and $p_{2}$ respectively and they
read
$$\begin{array}{l}
p_{1}=\left(E_{t}, \sqrt{E_{t}^{2}-m_{t}^{2}}sin\theta, 0,
\sqrt{E_{t}^{2}-m_{t}^{2}}cos\theta\right),\\
p_{2}=\left(E_{c}, -\sqrt{E_{c}^{2}-m_{c}^{2}}sin\theta, 0,
-\sqrt{E_{c}^{2}-m_{c}^{2}}cos\theta\right),
\end{array}
\eqno{(8)}
$$
where
$$
E_{t}=\frac{1}{2}\left(\sqrt{\hat{s}}+(m_{t}^{2}-m_{c}^{2})/\sqrt{\hat{s}}
\right),~~~~~
E_{c}=\frac{1}{2}\left(\sqrt{\hat{s}}-(m_{t}^{2}-m_{c}^{2})/\sqrt{\hat{s}}
\right).
\eqno{(9)}
$$
$p_{3}$ and $p_{4}$ are the four-momenta of the initial $\gamma$ and they
are
$$
p_{3}=\left(\frac{1}{2}\sqrt{\hat{s}}, 0, 0, \frac{1}{2}\sqrt{\hat{s}}\right),
~~~~
p_{4}=\left(\frac{1}{2}\sqrt{\hat{s}}, 0, 0, -\frac{1}{2}\sqrt{\hat{s}}\right).
\eqno{(10)}
$$

The corresponding matrix element for all the diagrams in figure 1(a) and
figure 1(b) is written as
$$
M = M^{\hat{s}}+M^{\hat{t}}+M^{\hat{u}}
\eqno{(11)}
$$

The upper indexes  $\hat{s},$ $\hat{t},$ and $\hat{u}$ represent the
amplitudes corresponding to the s-channel diagrams, t-channel and u-channel
diagrams in figure 1(a) and figure 1(b) respectively. The variables
$\hat{s}$, $\hat{t}$ and $\hat{u}$ are usual Mandelstam variables in the
center of mass system of $\gamma \gamma$. Their definitions are:
$$
\begin{array}{l}
\hat{s}=(p_{1}+p_{2})^{2}=(p_{3}+p_{4})^{2},~~~~
\hat{t}=(p_{1}-p_{3})^{2}=(p_{2}-p_{4})^{2},\\
\hat{u}=(p_{1}-p_{4})^{2}=(p_{2}-p_{3})^{2}.
\end{array}
\eqno{(12)}
$$

We collect all the explicit expressions of the amplitudes appearing in
equation (11) in the appendix. The total cross section for
$\gamma\gamma \rightarrow t\bar{c}+\bar{t}c$ can be written in the form
$$
\hat{\sigma}(\hat{s})=\frac{2 N_{c}}{16\pi \hat{s}^2}
\int_{\hat{t^{-}}}^{\hat{t^{+}}} d\hat{t} \vert \bar{M}\vert^{2}
\eqno{(13)}
$$
where $\vert \bar{M}\vert^{2}$ is the initial spin-averaged matrix element
squared, the color factor $N_{c}=3$ and
$\hat{t^{\pm}}=1/2(m_{t}^{2}+m_{c}^{2}-\hat{s})\pm \sqrt{E_{t}^{2}-m_{t}^{2}}
\sqrt{\hat{s}}$. The total cross section for $e^{+}e^{-}\rightarrow \gamma
\gamma\rightarrow t\bar{c}+\bar{t}c$ can be obtained by folding the
$\hat{\sigma}$, the cross section for
$\gamma\gamma\rightarrow t\bar{c}+\bar{t}c$, with the photon luminosity,
$$
\sigma (s)=\int^{x_{max}}_{(m_{t}+m_{c})/\sqrt{s}} dz \frac{dL_{\gamma\gamma}}
{dz}\hat{\sigma}~~~(\gamma\gamma\rightarrow t\bar{c}+\bar{t}c~at~\hat{s}=z^{2}s),
\eqno{(14)}
$$
where $\sqrt{s}$ and $\sqrt{\hat{s}}$ are the $e^{+}e^{-}$ and $\gamma\gamma$
CMS energies respectively, and the quantity $dL_{\gamma\gamma}/dz$ is the
photon luminosity, which is defined as
$$
\frac{dL_{\gamma\gamma}}{dz}=2z\int^{x_{max}}_{z^{2}/x_{max}}
\frac{dx}{x} F_{\gamma /e}(x)F_{\gamma /e}(z^{2}/x).
\eqno{(15)}
$$
For unpolarized initial electrons and laser photon beams, the energy spectrum
of the back-scattered photon is given by\cite{telnov}
$$
F_{\gamma /e}=\frac{1}{D(\xi)}\left[1-x+\frac{1}{1-x}-\frac{4x}{\xi(1-x)}
             +\frac{4x^{2}}{\xi^{2}(1-x)^{2}}\right],
\eqno{(16)}
$$
where
$$
D(\xi)=\left(1-\frac{4}{\xi}-\frac{8}{\xi^2}\right)ln(1+\xi)+\frac{1}{2}
        +\frac{8}{\xi}-\frac{1}{2(1+\xi)^{2}},
\eqno{(17)}
$$
and $\xi=4E_{0}\omega_{0}/m_{e}^{2}$. $m_{e}$ and $E_{0}$ are the mass and
energies of the incident electron, respectively. The dimensionless parameter $x$
represents the fraction of the energy of the incident electron carried by
the back-scattered photon. In our evaluation, we choose $\omega_{0}$ such that
it maximizes the back-scattered photon energy without spoiling the luminosity
by $e^{+}e^{-}$ pair production. Then we can get $\xi=2(1+\sqrt{2})\simeq
4.8$, $x_{max}\simeq 0.83$ and $D(\xi)\simeq 1.8$. That is a usual method
which was used in Ref.\cite{kcheung}.

{\Large{\bf III. Numerical Results and Discussion}}

In the numerical evolution we take the input parameters\cite{parameters} as
$m_{b}=4.5 GeV$, $m_{c}=1.35 GeV$, $m_{t}=175 GeV$, $M_{W}=80.2226 GeV$,
$G_{F}=1.166392\times 10^{-5} (GeV)^{-2}$ and $\alpha=1/137.036$.

  Figure 2 shows the cross sections for $\gamma\gamma\rightarrow t\bar{c}+
\bar{t}c$ as a function of the masses of the Higgs bosons $M_{s}$. The cross
sections are displayed for the three values of the $\gamma \gamma$ CMS energy
200 GeV, 400 GeV and 500 GeV respectively. Because there is no stringent bound
on the Higgs bosons masses, we choose $M_{s}$ in the range from 50 GeV to 800
GeV. The higher peak of each curve comes from s-channel resonance effects,
where $M_{s}=m_{h^{0}}=m_{A^{0}} \sim \sqrt{\hat{s}}$.
The smaller peak of each curve mainly comes from the contribution of the
quartic vertex diagram. For the curve $\sqrt{\hat{s}}\sim 200~GeV$, it is
located at about $\sqrt{\hat{s}} \sim 2~M_{s}$, whereas for
$\sqrt{\hat{s}} \sim 400~GeV$ and $500~GeV$, they are at $M_{s}\sim 150 GeV$.
From these curves we find that the cross section can be obviously enhanced
when $M_{s}$ gets close to $\sqrt{\hat{s}}$.

Figure 3 shows the cross sections of $\gamma\gamma\rightarrow t\bar{c}+
\bar{t}c$ as a function of $\sqrt{\hat{s}}$, and the three curves correspond
to the $M_{s}$ values 100 GeV, 250 GeV and 500 GeV respectively.
For $M_{s}=100 GeV$, the effects of the widths of the Higgs bosons are not
obvious, and s-channel resonance effects are suppressed, since
$\sqrt{\hat{s}}$ is far away from $M_{s}$. Therefore its cross section
goes down as $\sqrt{\hat{s}}$ increasing. When $\sqrt{\hat{s}}$
approaches the value of $M_{s}$, such as $M_{s}=250 GeV$, the cross section
will be enhanced by the s-channel resonance effects, and the width effects
become larger, since the $h^0, A^0 \rightarrow t\bar{c}+\bar{t}c$
channels are opened. The small peak of the dashed line, where
$\sqrt{\hat{s}}\sim 2 m_{t}=350~GeV$, comes from the contribution of the
neutral Higgs box diagrams. For $M_{s}=500 GeV$, the cross section
goes up from $\sqrt{\hat{s}}\simeq 300 GeV$ due to the s-channel resonance
effects and large width effects of $h^{0}$ and $A^{0}$.

In figure 4 we show the cross section of $e^{+}e^{-}\rightarrow\gamma\gamma
\rightarrow t\bar{c}+\bar{t}c$ as a function of center-of-mass energy of
electron-positron system $\sqrt{s}$. The cross section
may reach 0.11 femptobarn when $M_{s}=250~GeV$ and $\sqrt{s}=355~GeV$. For
$M_{s}=100 GeV$ the cross section is two orders smaller than that for
$M_{s}=250~GeV$. When $M_{s}=500~GeV$, the cross section is only of the
order $10^{-4}$ femptobarn. For a 500 GeV NLC operating in $e^{+}e^{-}$ mode
with 50 $fb^{-1}$ integrated luminosity, one can expect about
6 raw events when $M_{s}=250~GeV$. Since the cross section of this process
roughly scales as $\lambda^{4}$, if we let $\lambda \simeq 1$,
the cross section will be 4 times larger. That is to say about
24 raw events may be produced per year in unpolarized photon collisions.
That means the $t\bar{c}$ production may be detectable in future NLC experiment.
In Ref.\cite{houlin} Hou et al. pointed out that one could expect
$10^{2}\sim 10^{3}$ $\gamma \gamma \rightarrow h^0, A^0 \rightarrow
t\bar{c}+\bar{t}c$ raw events a year with 50 $fb^{-1}$ integrated luminosity,
where they assume that the on-mass-shell neutral Higgs bosons with
 $ 200~GeV < m_{h^{0},A^{0}} < 2~m_{t} \simeq 350~GeV$ are produced and the
photon polarizations have $<\lambda\lambda^{'}>\sim +1$.
This estimate doesn't contradict ours, if we make the same assumptions and
the parameters have their most favorable values.
For the process $e^{+}e^{-} \rightarrow t\bar{c}
+\bar{t}c$ , Atwood et al. got $R^{tc}\equiv  \frac{\sigma(e^{+}e^{-}
\rightarrow t\bar{c}+ \bar{t}c)}{\sigma(e^{+}e^{-} \rightarrow \gamma^{*}
\rightarrow \mu^{+} \mu^{-})} {}^{<}_{\sim}  few \times 10^{-5}$\cite{atwood},
when all the masses of Higgs bosons are degenerate, which amounts to less
than 0.1 event for a 500 GeV NLC with 50 $fb^{-1}$ integrated luminosity.
It shows clearly that with the same parameters, the process
$e^{+}e^{-} \rightarrow \gamma\gamma \rightarrow t\bar{c}+\bar{t}c $
occurs with larger cross section than $e^{+}e^{-}
\rightarrow t\bar{c}+\bar{t}c$ process.

In summary, from our one-loop calculation, we can conclude that it is
possible in the NLC that the process
$e^{+}e^{-} \rightarrow \gamma\gamma \rightarrow t\bar{c}+\bar{t}c$ can be
used to probe the flavor changing interactions in the context of THDM III
with clean signals. The NLC operating in photon-photon mode can produce
more events for discovering flavor changing scalar interactions than
in electron-positron collision mode.

One of the authors, Ma Wen-Gan, would like to thank the Institute of
Theoretical Physics of the University of Vienna and the Institute of High
Energy Physics of the Austrian Academy of Sciences for warm
hospitality extended to him during his stay under agreement of
the exchange program(Project number IV.B.12).

\vskip 5mm
\noindent
{\Large{\bf Appendix}}
\vskip 5mm

We adopt the same definitions of one-loop A, B, C and D integral functions as
in Ref.\cite{abcd} and the references therein. The dimension $D=4- \epsilon$.
The integral functions are defined as
$$
A_{0}(m)=-\frac{(2\pi\mu)^{4-D}}{i\pi^{2}} \int d^{D}q \frac{1}{[q^2-m^2]} ,
$$
$$
\{B_{1};B_{\mu};B_{\mu\nu}\}(p,m_1,m_2) =
\frac{(2\pi\mu)^{4-D}}{i\pi^{2}} \int d^{D}q
\frac{\{1;q_{\mu};q_{\mu\nu}\}}{[q^2-m_{1}^{2}][(q+p)^2-m_{2}^{2}]} ,
$$
$$
\{C_{0};C_{\mu};C_{\mu\nu};C_{\mu\nu\rho}\}(p_1,p_2,m_1,m_2,m_3) =
-\frac{(2\pi\mu)^{4-D}}{i\pi^{2}}
$$
$$
\times \int d^{D}q
\frac{\{1;q_{\mu};q_{\mu\nu};q_{\mu\nu\rho}\}}
{[q^2-m_{1}^{2}][(q+p_1)^2-m_{2}^{2}][(q+p_1+p_2)^2-m_{3}^{2}]} ,
$$
$$
\{D_{0};D_{\mu};D_{\mu\nu};D_{\mu\nu\rho};D_{\mu\nu\rho\alpha}\}
(p_1,p_2,p_3,m_1,m_2,m_3,m_4) =
\frac{(2\pi\mu)^{4-D}}{i\pi^{2}}
$$
$$
\times \int d^{D}q \{1;q_{\mu};q_{\mu\nu};q_{\mu\nu\rho};
q_{\mu\nu\rho\alpha}\}
$$
$$
\times \{[q^2-m_{1}^{2}][(q+p_1)^2-m_{2}^{2}][(q+p_1+p_2)^2-m_{3}^{2}]
[(q+p_1+p_2+p_3)^2-m_{4}^{2}]\}^{-1}.
$$

In our calculation we take the strange quark mass $m_{s}=0$.
The $M^{\hat{s}}$ in equation (11) can be written as
$$
\begin{array}{lll}
M^{\hat{s}}&=&
\frac{ig^{2}e^{2}}{36 \pi^{2} M_{W}^{2}} m_{t}\sqrt{m_{t}m_{c}}
\epsilon_{\mu}(p_{3})\epsilon_{\nu}(p_{4})
\bar{u}(p_{1})\\
&&\cdot\{
2 a_{h^{0}} m_{t}^2 (C_{0} + 4 C_{22} - 4 C_{23})[p_{3}, -p_{1}
 - p_{2}, m_{t}, m_{t}, m_{t}](p_{1}^{\mu}p_{1}^{\nu}+p_{1}^{\mu}p_{2}^{\nu}+
 p_{1}^{\nu}p_{2}^{\mu}+p_{2}^{\mu}p_{2}^{\nu})\\
&& + 2a_{h^{0}}m_{t}^2 (B_{0}[-p_{1} - p_{2}, m_{t}, m_{t}] -
       ((p_{1}+p_{2}) \cdot p_{3} C_{0} \\
&&       +4 C_{24})
        )[p_{3}, -p_{1} - p_{2}, m_{t}, m_{t}, m_{t}]
+ 9[ m_{b}^2 C_{0} +m_{t} m_{c} (C_{11} - C_{12}) + m_{t}^2 C_{12}\\
&&- \gamma_{5}(m_{b}^2 C_{0} - m_{t} m_{c} (C_{11} - C_{12}) - m_{t}^2 C_{12})]
  [p_{2}, -p_{1} - p_{2}, m_{b}, M_{s}, M_{s}]
        )g^{\mu\nu}\\
&&      + 2 i a_{A^{0}} m_{t}^2 C_{0}[p_{3}, -p_{1} - p_{2}, m_{t}, m_{t}, m_{t}]
      \epsilon^{\mu\nu\alpha\beta}\gamma_{5}
        (p_{1}^{\alpha}p_{3}^{\beta}+p_{2}^{\alpha}p_{3}^{\beta})
        \}v(p_{2}),
\end{array}
\eqno{(A.1)}
  $$
where
$$a_{h^{0}}=\frac{1}{\hat{s}-m_{h^{0}}^2+i\Gamma_{h^{0}}m_{h^{0}}},~~~~~
  a_{A^{0}}=\frac{1}{\hat{s}-m_{A^{0}}^2+i\Gamma_{A^{0}}m_{A^{0}}}.$$

The amplitude of $M^{\hat{t}}$ can be written as
$$
\begin{array}{lll}
M^{\hat{t}}&=&
\frac{ig^{2}e^{2}}{288 \pi^{2} M_{W}^{2}} \sqrt{m_{t}m_{c}}
\epsilon_{\mu}(p_{3})\epsilon_{\nu}(p_{4}) \bar{u}(p_{1})
 (f_{1}p_{1}^{\mu}p_{1}^{\nu}+f_{2}p_{1}^{\mu}p_{2}^{\nu}+
f_{3}p_{1}^{\nu}p_{2}^{\mu}
+f_{4}p_{2}^{\mu}p_{2}^{\nu}\\
&&+ f_{5}\gamma^{\nu}p_{1}^{\mu}+f_{6}\gamma^{\mu}p_{1}^{\nu}
+ f_{7}\gamma^{\nu}p_{2}^{\mu}
+f_{8}\gamma^{\mu}p_{2}^{\nu}
+ f_{9}\gamma^{\mu}\gamma^{\nu}+f_{10}\gamma^{\nu}\gamma^{\mu}
+ f_{11}\rlap/p_{3}p_{1}^{\mu}p_{1}^{\nu}\\
&&+f_{12}\rlap/p_{3}p_{1}^{\mu}p_{2}^{\nu}
+ f_{13}\rlap/p_{3}p_{1}^{\nu}p_{2}^{\mu}
+f_{14}\rlap/p_{3}p_{2}^{\mu}p_{2}^{\nu}
+ f_{15}\rlap/p_{3}\gamma^{\nu}p_{1}^{\mu}
+f_{16}\rlap/p_{3}\gamma^{\mu}p_{1}^{\nu}
+ f_{17}\rlap/p_{3}\gamma^{\nu}p_{2}^{\mu}\\
&&+f_{18}\rlap/p_{3}\gamma^{\mu}p_{2}^{\nu}
+ f_{19}\rlap/p_{3}\gamma^{\mu}\gamma^{\nu}
+f_{20}\rlap/p_{3}\gamma^{\nu}\gamma^{\mu}
+ f'_{1}\gamma_{5} p_{1}^{\mu}p_{1}^{\nu}
+f'_{2}\gamma_{5} p_{1}^{\mu}p_{2}^{\nu}+
f'_{3}\gamma_{5} p_{1}^{\nu}p_{2}^{\mu}\\
&&+f'_{4}\gamma_{5} p_{2}^{\mu}p_{2}^{\nu}+
f'_{5}\gamma_{5} \gamma^{\nu}p_{1}^{\mu}
+f'_{6}\gamma_{5} \gamma^{\mu}p_{1}^{\nu}
+ f'_{7}\gamma_{5} \gamma^{\nu}p_{2}^{\mu}
+f'_{8}\gamma_{5} \gamma^{\mu}p_{2}^{\nu}
+ f'_{9}\gamma_{5} \gamma^{\mu}\gamma^{\nu}\\
&&+f'_{10}\gamma_{5} \gamma^{\nu}\gamma^{\mu}
+ f'_{11}\gamma_{5} \rlap/p_{3}p_{1}^{\mu}p_{1}^{\nu}
+f'_{12}\gamma_{5} \rlap/p_{3}p_{1}^{\mu}p_{2}^{\nu}
+ f'_{13}\gamma_{5} \rlap/p_{3}p_{1}^{\nu}p_{2}^{\mu}
+f'_{14}\gamma_{5} \rlap/p_{3}p_{2}^{\mu}p_{2}^{\nu}\\
&&+ f'_{15}\gamma_{5} \rlap/p_{3}\gamma^{\nu}p_{1}^{\mu}
+f'_{16}\gamma_{5} \rlap/p_{3}\gamma^{\mu}p_{1}^{\nu}
+ f'_{17}\gamma_{5} \rlap/p_{3}\gamma^{\nu}p_{2}^{\mu}\\
&&+f'_{18}\gamma_{5} \rlap/p_{3}\gamma^{\mu}p_{2}^{\nu}
+ f'_{19}\gamma_{5} \rlap/p_{3}\gamma^{\mu}\gamma^{\nu}
+f'_{20}\gamma_{5} \rlap/p_{3}\gamma^{\nu}\gamma^{\mu} )v(p_{2}),
\end{array}
\eqno{(A.2)}
$$

where the $f_{i}s$ and $f'_{i}s$ are expressed explicitly as,
$$
\begin{array}{lll}
f'_{1}& =&
 -36 (
m_{b}^2 D_{13} +m^{2}_{t} (D_{26}+D_{38})\\
&&-m_{t} m_{c} (D_{25}+D_{310})
   )[p_{2}, -p_{4}, -p_{3}, m_{b}, M_{s}, M_{s}, M_{s}]\\
&&+12 (
m_{b}^2 (D_{11}-D_{12})
+m_{t} m_{c} (D_{25}-D_{26}-D_{310}+D_{35})\\
&&+m_{t}^2 (D_{21}-D_{24}+D_{31}-D_{34})
   )[-p_{1}, p_{3}, -p_{2}, m_{b}, M_{s}, M_{s}, m_{b}]\\
&& -4 (
m_{b}^2 (D_{22}-D_{24})-m_{t} m_{c} (D_{310}-D_{38})\\
&&+m_{t}^2 (D_{32}-D_{36})
   )[p_{3}, -p_{1}, -p_{2}, m_{b}, m_{b}, M_{s}, m_{b}],
\end{array}
\eqno{(A.3)}
$$

$$
\begin{array}{lll}
f_{1}& =&
f'_{1}(m_{i}^2\rightarrow -m_{i}^2, i=b,t)
+ 32 m_{t}^2 (D_{11} - D_{12} + 2 D_{21} - 2 D_{24}\\
&&      - 2 D_{25} + 2 D_{26} + D_{31} + 2 D_{310} - D_{34} - 2 D_{35}\\
&&        + D_{37} - D_{39})
        [p_{1}, -p_{3}, -p_{4}, M_{s}, m_{t}, m_{t}, m_{t}],
\end{array}
\eqno{(A.4)}
$$

$$
\begin{array}{lll}
f'_{2}& =&
 24 a_{1} (
   m_{b}^2 C_{0} - m_{t} m_{c} (C_{11}+C_{21})
   + m_{t}^2 (C_{12}+C_{23}) ) [p_{2}, -p_{4}, m_{b}, M_{s}, M_{s}]\\
&&- 24 a_{1} ( m_{b}^2 C_{0} - m_{t} m_{c} C_{22}
   +m_{t}^2 (C_{12}+C_{23})
   )[-p_{4}, p_{2}, m_{b}, m_{b}, M_{s}]\\
&& -36 (
 m_{b}^2 (D_{0} -D_{13})
 +m_{t} m_{c} (D_{11}+D_{21}-D_{25}-D_{310}+D_{34})  \\
&& +m_{t}^2 (D_{12}+D_{22}+ D_{24}-D_{26}+D_{36}-D_{38})
   )[p_{2}, -p_{4}, -p_{3}, m_{b}, M_{s}, M_{s}, M_{s}]\\
&&-12 ( m_{b}^2 (D_{0}+D_{12})
- m_{t} m_{c} (D_{23}-D_{26}-D_{310}+D_{37})
+m_{t}^2 (D_{11}-D_{13}+D_{21}\\
&&+D_{24}-2 D_{25}+D_{34} +D_{35})
   )[-p_{1}, p_{3}, -p_{2}, m_{b}, M_{s}, M_{s}, m_{b}]\\
&& +4 (
m_{b}^2 (D_{12}+D_{24})+m_{t} m_{c} (D_{310}-D_{39})\\
&&+m_{t}^2 (D_{22}-D_{26}+D_{36}-D_{38})
   )[p_{3}, -p_{1}, -p_{2}, m_{b}, m_{b}, M_{s}, m_{b}],
\end{array}
\eqno{(A.5)}
$$

$$
\begin{array}{lll}
f_{2}& =&
f'_{2}(m_{i}^2\rightarrow -m_{i}^2, i=b,t)
+ 32 a_{1}m_{t}  ( m_{c} (C_{11} + C_{21}) \\
&&+m_{t} (C_{12} + C_{23})
        )[-p_{2}, p_{4}, M_{s}, m_{t}, m_{t}]
- 32 m_{t}^2
      (D_{12} - D_{13} + D_{23} + 2 D_{24}\\
&&      - D_{25} - 2 D_{26} - 2 D_{310} + D_{34} + D_{39})
        [p_{1}, -p_{3}, -p_{4}, M_{s}, m_{t}, m_{t}, m_{t}],
\end{array}
\eqno{(A.6)}
$$

$$
\begin{array}{lll}
f'_{3} &=&
 36 (
    m_{b}^2 (D_{25}-D_{26}) +m_{t} m_{c} D_{35}\\
&&    -m_{t}^2 (D_{310}-D_{38})
   )[p_{2}, -p_{4}, -p_{3}, m_{b}, M_{s}, M_{s}, M_{s}]
+12 (
m_{b}^2 (D_{25}-D_{26})\\
&&+m_{t} m_{c} (D_{37}-D_{39})-m_{t}^2 (D_{310}-D_{35})
   )[-p_{1}, p_{3}, -p_{2}, m_{b}, M_{s}, M_{s}, m_{b}]\\
&& +4 (
m_{b}^2 (D_{25}-D_{26})+m_{t} m_{c} (D_{37}-D_{39})\\
&&+m_{t}^2 (D_{310}-D_{38})
   )[p_{3}, -p_{1}, -p_{2}, m_{b}, m_{b}, M_{s}, m_{b}],
\end{array}
\eqno{(A.7)}
$$

$$
\begin{array}{lll}
f_{3} &=&
f'_{3}(m_{i}^2\rightarrow -m_{i}^2, i=b,t)\\
&&+ 32m_{t}^2 (D_{37} - D_{39} -D_{25} + D_{26}
 + D_{310} - D_{35})[p_{1}, -p_{3}, -p_{4}, M_{s}, m_{t}, m_{t}, m_{t}],
\end{array}
\eqno{(A.8)}
$$

$$
\begin{array}{lll}
f'_{4}& =&
 -36 (
m_{b}^2 (D_{11}-D_{12}+D_{26})
-m_{t} m_{c} (D_{21} +D_{31} -D_{35})
-m_{t}^2 (D_{22}-D_{24}\\
&&+D_{310}-D_{34}
+D_{36}-D_{38}) )[p_{2}, -p_{4}, -p_{3}, m_{b}, M_{s}, M_{s}, M_{s}]\\
&&-12 (
m_{b}^2 (D_{13}+D_{26})
-m_{t} m_{c} (D_{33}-D_{39})-m_{t}^2 (D_{23}-D_{25}\\
&&-D_{310}+D_{37})
   )[-p_{1}, p_{3}, -p_{2}, m_{b}, M_{s}, M_{s}, m_{b}]\\
&& +4 (
m_{b}^2 (D_{13}+D_{25})-m_{t} m_{c} (D_{33}-D_{37})\\
&&-m_{t}^2 (D_{23}-D_{26}-D_{310}+D_{39})
   )[p_{3}, -p_{1}, -p_{2}, m_{b}, m_{b}, M_{s}, m_{b}],
\end{array}
\eqno{(A.9)}
$$

$$
f_{4} =
f'_{4}(m_{i}^2\rightarrow -m_{i}^2, i=b,t)
+ 32m^{2}_{t} ( D_{26} + D_{310} - D_{23} - D_{39}
        )[p_{1}, -p_{3}, -p_{4}, M_{s}, m_{t}, m_{t}, m_{t}],
\eqno{(A.10)}
$$

$$
\begin{array}{lll}
f'_{5}& =&
 8 a_{2} a_{3}  m_{t} (m_{b}^2 B_{0} + m_{t}^2 B_{1})[-p_{1}, m_{b}, M_{s}]\\
&& +8 a_{1} a_{3} m_{t} (m_{b}^2 B_{0} - m_{t} m_{c} B_{1})[p_{2}, m_{b}, M_{s}]\\
&&- 4 a_{1} a_{2} (
   m_{t} (m_{b}^2 B_{0} + m_{t}^2 B_{1}) + 2 m_{t} (p_{1}\cdot p_{3}) B_{1}
   )[-p_{1} + p_{3}, m_{t}, M_{s}] \\
&& +12 a_{2} m_{t} (
  (m_{b}^2 C_{0} + m_{t}^2 (C_{11}+C_{21}))\\
&& + 2  (C_{24}+(p_{1}\cdot p_{3}) (C_{12}+C_{23}))
   )[-p_{1}, p_{3}, m_{b}, M_{s}, M_{s}]\\
&& -24 a_{1} m_{t} C_{24}[p_{2}, -p_{4}, m_{b}, M_{s}, M_{s}]
 - 12 a_{2} m_{t} (2 C_{24}+m_{b}^2 (C_{0}+C_{12})\\
&&  -m_{c} m_{t} C_{22})
   [p_{3}, -p_{1}, m_{b}, m_{b}, M_{s}]
 + 12 a_{1}m_{t} (
2 m_{b}^2  C_{0} - 2 C_{24}\\
&&+m_{c}^2 (C_{22}-C_{23})-m_{t} m_{c} C_{12}
        +m_{t}^2 (C_{11}+C_{21}))\\
&&+ 2 (p_{1}\cdot p_{2}-p_{2}\cdot p_{3}) (C_{11}-C_{12}+C_{21}-C_{23})\\
&&-2 p_{1}\cdot p_{3} (C_{11}+C_{21})
   )[-p_{4}, p_{2}, m_{b}, m_{b}, M_{s}]\\
&& -36 m_{t} (D_{27}+D_{312}) [p_{2}, -p_{4}, -p_{3}, m_{b}, M_{s}, M_{s}, M_{s}]
+6 (
 m_{t} (4 D_{27}+4 D_{311}\\
&&+m_{c}^2 (D_{23}+D_{37})
            +m_{t} m_{c} (D_{13}+D_{25})
            +m_{t}^2 (D_{11}-D_{21}\\
&&            -D_{31})) - 2 m_{t} ((p_{1}\cdot p_{2}) (D_{13}+2 D_{25}+D_{35})
              +(p_{1}\cdot p_{3}) (D_{12}+D_{24}+D_{34})\\
&&              +(p_{2}\cdot p_{3}) (D_{13}+D_{25}+D_{26}+D_{310}))
   )[-p_{1}, p_{3}, -p_{2}, m_{b}, M_{s}, M_{s}, m_{b}]\\
&& -2 (
m_{t} (2 D_{27}+2 D_{312}+m_{b}^2 D_{0}
-m_{c}^2 D_{23}-m_{t} m_{c} (D_{22}-D_{26}))\\
&&+m_{b}^2 m_{t} D_{12} +2 m_{t} (p_{2}\cdot p_{3}) (D_{25}-D_{26})
   )[p_{3}, -p_{1}, -p_{2}, m_{b}, m_{b}, M_{s}, m_{b}],
\end{array}
\eqno{(A.11)}
$$

$$
\begin{array}{lll}
f_{5}& =&
f'_{5}( m_{t}m_{c}\rightarrow -m_{t}m_{c})
 -16  a_{2} a_{3} m_{t}^3 B_{1}[-p_{1}, m_{t}, M_{s}]
+ 16  a_{1} a_{2}m_{t} (2 p_{1} \cdot p_{3}\\
&&- m^{2}_{t}) B_{1}[-p_{1} + p_{3}, m_{t}, M_{s}]
+16  a_{1} a_{3} m^{2}_{t}m_{c} B_{1}[p_{2}, m_{t}, M_{s}]\\
&&-16 a_{2} m_{t} ( m^{2}_{t} C_{0} + m_{t} m_{c}(2 C_{11} + C_{21})\\
&&+2 C_{24}
        )[p_{1}, -p_{3}, M_{s}, m_{t}, m_{t}]
- 16 a_{1}m_{t}  ( 2 C_{24} - m_{c}^2 ( \\
&& +C_{21} - 2 C_{23}) +
       m_{c} m_{t}  C_{11} - m_{t}^2 ( C_{12} + C_{22}- C_{0}) -
        2 (p_{1} \cdot p_{2} \\
&&        - p_{2} \cdot p_{3}) (C_{22} - C_{23}) - 2 p_{1} \cdot p_{3} (C_{12} + C_{22})
        )[-p_{2}, p_{4}, M_{s}, m_{t}, m_{t}]        \\
&&- 16m_{t} ( 2 D_{311} - 2 D_{313}  -
      m_{t}m_{c} (D_{21}-D_{25}) - m_{t}^2 (D_{0} + 2 D_{11}-D_{13})  \\
&&      - 2 (p_{2} \cdot p_{3}) (D_{25} - D_{26})
        )[p_{1}, -p_{3}, -p_{4}, M_{s}, m_{t}, m_{t}, m_{t}],
\end{array}
\eqno{(A.12)}
$$

$$
\begin{array}{lll}
f'_{6}& =&
 -36 m_{t} D_{313}[p_{2}, -p_{4}, -p_{3}, m_{b}, M_{s}, M_{s}, M_{s}]
+12 m_{t} (D_{311}\\
&&-D_{312})[-p_{1}, p_{3}, -p_{2}, m_{b}, M_{s}, M_{s}, m_{b}]
 -2 ( m_{t} (4 D_{311}-4 D_{312}+m_{b}^2 (D_{11}-D_{12})\\
&&-m_{c}^2 (D_{37}-D_{39})+m_{t}^2 (D_{32}-D_{36}))
-2 m_{t} ( p_{1}\cdot p_{2} (D_{310}-D_{38})
          +p_{1}\cdot p_{3} (D_{22}-D_{24}\\
&&          -D_{34}+D_{36})
          +p_{2}\cdot p_{3} (D_{310}-D_{35}))
   )[p_{3}, -p_{1}, -p_{2}, m_{b}, m_{b}, M_{s}, m_{b}],
\end{array}
\eqno{(A.13)}
$$

$$
\begin{array}{lll}
f_{6}& =& f'_{6}+
16 m_{t}( 4 D_{311} - 4 D_{312} -
       m_{t}^2 (2 D_{21}  - 2 D_{24} - 2 D_{25}  + 2 D_{26}
       + D_{31} + 2 D_{310} \\
&&- D_{34}  - 2 D_{35} + D_{37} - D_{39}) +
        2 (p_{1} \cdot p_{2}) (D_{25} - D_{26} - D_{310} + D_{35} -
        D_{37}+ D_{39}) \\
&&        - 2 (p_{1} \cdot p_{3}) (D_{22} - D_{24} + D_{25} - D_{26} -
        D_{34} + D_{35} + D_{36} - D_{37} - D_{38} + D_{39})\\
&&        - 2 (p_{2} \cdot p_{3}) (D_{25} - D_{26}
        + D_{310} - D_{37} - D_{38} + D_{39})
        )[p_{1}, -p_{3}, -p_{4}, M_{s}, m_{t}, m_{t}, m_{t}],
\end{array}
\eqno{(A.14)}
$$

$$
\begin{array}{lll}
f'_{7}& = &
 -36 m_{t} (D_{312}-D_{311})[p_{2}, -p_{4}, -p_{3}, m_{b}, M_{s}, M_{s}, M_{s}]
-6 (
m_{t} (4 D_{313}\\
&&-m_{c}^2 D_{33}-m_{t} m_{c} D_{23}
-m_{t}^2 (D_{25}+D_{35}))-m_{b}^2 m_{c} D_{13} \\
&&-2 m_{t} (p_{1}\cdot p_{2} (D_{23}+D_{37})
              -p_{1}\cdot p_{3} (D_{26}+D_{310})\\
&&              -p_{2}\cdot p_{3} (D_{23}+D_{39}))
   )[-p_{1}, p_{3}, -p_{2}, m_{b}, M_{s}, M_{s}, m_{b}]\\
&& +2 (
m_{t} (4 D_{313}-m_{c}^2 D_{33}
-m_{t} m_{c} D_{23}-m_{t}^2 (D_{26}+D_{38}))-m_{b}^2 m_{c} D_{13}\\
&& +2 m_{t} (p_{1}\cdot p_{2} (D_{23}+D_{39})+p_{1}\cdot p_{3} (D_{25}+D_{310})\\
&&              +p_{2}\cdot p_{3} (D_{23}+D_{37}))
   )[p_{3}, -p_{1}, -p_{2}, m_{b}, m_{b}, M_{s}, m_{b}],
\end{array}
\eqno{(A.15)}
$$

$$
\begin{array}{lll}
f_{7}& = &
f'_{7}( m_{t}m_{c}\rightarrow -m_{t}m_{c})
+16 m_{t}( m_{t}^2 (-D_{13} - D_{23}
        + D_{25} + D_{33} + D_{35} - 2 D_{37})\\
&&        -4 D_{313} + 2 (p_{1} \cdot p_{2}) (D_{33} - D_{37}) +
        2 (p_{1} \cdot p_{3}) (D_{23} - D_{25} - D_{310} - D_{33} \\
&&        + D_{37} + D_{39}) -
        2 (p_{2} \cdot p_{3}) (D_{33} - D_{39})
        )[p_{1}, -p_{3}, -p_{4}, M_{s}, m_{t}, m_{t}, m_{t}],
\end{array}
\eqno{(A.16)}
$$

$$
\begin{array}{lll}
f'_{8} &=&
 24 a_{1} m_{t} (p_{1}\cdot p_{3}) (C_{12}+C_{23})[p_{2}, -p_{4}, m_{b}, M_{s}, M_{s}]\\
&& -24 a_{1} m_{t} (p_{1}\cdot p_{3}) (C_{12}+C_{23})[-p_{4}, p_{2}, m_{b}, m_{b}, M_{s}]\\
&& -36 m_{t} (D_{313}-D_{27}-D_{311})[p_{2}, -p_{4}, -p_{3}, m_{b}, M_{s}, M_{s}, M_{s}]\\
&&-12 m_{t} (D_{27}+D_{312}-D_{313})[-p_{1}, p_{3}, -p_{2}, m_{b}, M_{s}, M_{s}, m_{b}]\\
&& -2 (
m_{t} (2 D_{27}+4 D_{311}+2 D_{313}+m_{b}^2 (D_{0}+D_{11})-m_{c}^2 (D_{23}+D_{37})
-m_{t}^2 (D_{22}+D_{36})\\
&&-2 m_{t} (p_{1}\cdot p_{2} (D_{26}+D_{310})
         -p_{1}\cdot p_{3} (D_{12}-D_{13}+2 D_{24}-D_{26}+D_{34})\\
&&         -p_{2}\cdot p_{3} (D_{25}+D_{35}))
   )[p_{3}, -p_{1}, -p_{2}, m_{b}, m_{b}, M_{s}, m_{b}],
\end{array}
\eqno{(A.17)}
$$

$$
\begin{array}{lll}
f_{8} &=&  f'_{8}
-32 a_{1}m_{t} (p_{1} \cdot p_{3}) C_{12}[-p_{2}, p_{4}, M_{s}, m_{t}, m_{t}]
+ 16m_{t} (6 D_{313} -2 D_{27}\\
&&- 4 D_{312}
+ m_{t}^2 (2 D_{23}+ 2 D_{24} - 2 D_{25} - 2 D_{26} - 2 D_{310} \\
&&      - D_{33} + D_{34} - D_{35} + 2 D_{37}+D_{39}) +
        2 (p_{1} \cdot p_{2}) (D_{23} - D_{26} - D_{310} - D_{33} + D_{37}  +
        D_{39})\\
&&        + 2 (p_{1} \cdot p_{3}) (D_{25}-D_{22} - 2 D_{23}  + 2 D_{26} +
        2 D_{310} +
        D_{33} - D_{36} - D_{37} + D_{38}-2 D_{39}) \\
&&        + 2 (p_{2} \cdot p_{3}) (  D_{26}-D_{23} + D_{33}
+ D_{38} - 2 D_{39})
        )[p_{1}, -p_{3}, -p_{4}, M_{s}, m_{t}, m_{t}, m_{t}] ,
\end{array}
\eqno{(A.18)}
$$

$$
\begin{array}{lll}
f'_{9} &=&
 4a_{2} a_{3} m_{t}^2 (m_{b}^2 B_{0} - m_{t}^2 B_{1})[-p_{1}, m_{b}, M_{s}]\\
&&+ 4 a_{1} a_{2} (p_{1}\cdot p_{3}) (m_{t} m_{c} B_{1}-m_{b}^2 B_{0})[-p_{1} + p_{3}, m_{b}, M_{s}]\\
&& -12 a_{2} m_{t}^2 C_{24}[-p_{1}, p_{3}, m_{b}, M_{s}, M_{s}]
 -6 a_{2} m_{t}^2 (2 C_{24}+ m_{b}^2 C_{0}-m_{t}^2 C_{22})\\
&&- 2 (p_{1}\cdot p_{3}) (m_{b}^2 C_{0} - m_{c}^2 C_{12}-m_{t}^2 C_{23})
   )[p_{3}, -p_{1}, m_{b}, m_{b}, M_{s}]\\
&& -12 a_{1} m_{t} m_{c} (p_{1}\cdot p_{3}) C_{0}[-p_{4}, p_{2}, m_{b}, m_{b}, M_{s}]\\
&& +18 (
m_{b}^2 D_{27} - m_{t} m_{c} D_{311} + m_{t}^2 D_{312}
   )[p_{2}, -p_{4}, -p_{3}, m_{b}, M_{s}, M_{s}, M_{s}]\\
&&-6 (
 m_{t} m_{c} D_{313}+m_{t}^2 (D_{27}+D_{311})
   )[-p_{1}, p_{3}, -p_{2}, m_{b}, M_{s}, M_{s}, m_{b}]\\
&&-(2 m_{b}^2 D_{27}+m_{b}^4 D_{0}-m_{b}^2 m_{c}^2 D_{23}-m_{b}^2 m_{t}^2 D_{22}
+m_{t} m_{c} (2 D_{27}+4 D_{313}+m_{b}^2 (D_{0}+D_{13})\\
&&-m_{c}^2 (D_{23}+D_{33})-m_{t}^2 (D_{22}+D_{38}))
+m_{t}^2 (2 D_{27}+4 D_{312}+m_{b}^2 (D_{0}+D_{12})\\
&&-m_{c}^2 (D_{23}+D_{39})-m_{t}^2 (D_{22}+D_{32}))
+2 m_{b}^2 (p_{1}\cdot p_{2} D_{26}
-p_{1}\cdot p_{3} (D_{0}-D_{24}) +p_{2}\cdot p_{3} D_{25})\\
&&-2 (p_{1}\cdot p_{2} (m_{t}m_{c} D_{39}+m_{t}^2 (D_{26}+D_{38}))
-p_{1}\cdot p_{3} (m_{t}m_{c} (D_{12}-D_{13}+D_{310})\\
&&+m_{t}^2 (D_{24}+D_{36}))
-p_{2}\cdot p_{3} (m_{t}m_{c} D_{37}+m_{t}^2 (D_{25}+D_{310})))
   )[p_{3}, -p_{1}, -p_{2}, m_{b}, m_{b}, M_{s}, m_{b}],
\end{array}
\eqno{(A.19)}
$$

$$
\begin{array}{lll}
f_{9} &=&
- f'_{9}(m_{t} m_{c}\rightarrow -m_{t} m_{c})\\
&&+8 a_{2} a_{3} m_{t}^4 B_{1}[-p_{1}, m_{t}, M_{s}]
+ 16  a_{1} a_{2} m_{t} m_{c}(p_{1} \cdot p_{3}) B_{1}[-p_{1} + p_{3}, m_{t}, M_{s}]\\
&&+ 8 a_{2} m_{t}^2 ( 2 C_{24} + m_{t}^2 (2 C_{11}+C_{21}) +
    2 (p_{1} \cdot p_{3}) (C_{0} + C_{11} \\
&&    +C_{12}+C_{23})
        )[p_{1}, -p_{3}, M_{s}, m_{t}, m_{t}]
+ 16 a_{1}  (p_{1} \cdot p_{3}) m_{t}m_{c} (C_{0}\\
&&+ C_{11})[-p_{2}, p_{4}, M_{s}, m_{t}, m_{t}]
+4m_{t} (
 4 m_{t} (D_{27} + 2 D_{311} - 2 D_{313})\\
&&  - 4 m_{c} m_{t}^2 (  D_{11} -  D_{13})
  - m_{t}^3 ( 4 D_{21} +  4 D_{23} -
        8 D_{25} + 2 D_{31} - 2 D_{33}\\
&&        - 6 D_{35}  + 6 D_{37})
        + 4 (p_{1} \cdot p_{2})  (m_{c} D_{13} -
        m_{t} (D_{23} - D_{25} - D_{33}\\
&&        - D_{35} + 2 D_{37}))
        + 4 (p_{1} \cdot p_{3}) (m_{c} D_{12}
        + m_{t} (D_{11} - D_{13}\\
&&        + D_{21} + 2 D_{23}  + D_{24}
        - 3 D_{25}  - D_{26} - 2D_{310} - D_{33} +D_{34}  - D_{35}\\
&&      + 2 D_{37} + D_{39})) -
        4 (p_{2} \cdot p_{3}) (m_{c} D_{13}
        - m_{t} (D_{23}-D_{25}\\
&&        -D_{26}-D_{310}-D_{33}+ D_{37} + D_{39}))
        )[p_{1}, -p_{3}, -p_{4}, M_{s}, m_{t}, m_{t}, m_{t}],
\end{array}
\eqno{(A.20)}
$$

$$
\begin{array}{lll}
f'_{10}& =&
 18 (
m_{b}^2 D_{27} - m_{t} m_{c} D_{311}+m_{t}^2 D_{312}
   )[p_{2}, -p_{4}, -p_{3}, m_{b}, M_{s}, M_{s}, M_{s}]\\
&&-6 (
m_{b}^2 D_{27}+m_{t} m_{c} D_{313}+m_{t}^2 D_{311}
   )[-p_{1}, p_{3}, -p_{2}, m_{b}, M_{s}, M_{s}, m_{b}]\\
&&+ 2 (m_{b}^2 D_{27} +m_{t} m_{c} D_{313}+m_{t}^2 D_{312}
   )[p_{3}, -p_{1}, -p_{2}, m_{b}, m_{b}, M_{s}, m_{b}],
\end{array}
\eqno{(A.21)}
$$

$$
\begin{array}{lll}
f_{10}& =&
f'_{10}(m_{i}^2\rightarrow -m_{i}^2,~i=b,t)\\
&&-16m_{t}^2 (D_{27} + D_{311} - D_{313})
         [p_{1}, -p_{3}, -p_{4}, M_{s}, m_{t}, m_{t}, m_{t}],
\end{array}
\eqno{(A.22)}
$$

$$
\begin{array}{lll}
f'_{11}& = &
 36 m_{t} (D_{23}-D_{26}-D_{38}+D_{39})[p_{2}, -p_{4}, -p_{3}, m_{b}, M_{s}, M_{s}, M_{s}]\\
&&-12 m_{t} (D_{22}-D_{24}-D_{34}+D_{36})[-p_{1}, p_{3}, -p_{2}, m_{b}, M_{s}, M_{s}, m_{b}]\\
&& -4 m_{t} (D_{22}-D_{24}-D_{34}+D_{36})[p_{3}, -p_{1}, -p_{2}, m_{b}, m_{b}, M_{s}, m_{b}],
\end{array}
\eqno{(A.23)}
$$

$$
\begin{array}{lll}
f_{11}& = &f'_{11}+
32 m_{t}( D_{22} - D_{24} + D_{25} - D_{26} - D_{34} + D_{35} +
         D_{36} - D_{37}\\
&&         - D_{38} + D_{39}
         )[p_{1}, -p_{3}, -p_{4}, M_{s}, m_{t}, m_{t}, m_{t}],
\end{array}
\eqno{(A.24)}
$$

$$
\begin{array}{lll}
f'_{12}& = &
 36 m_{t} (D_{12}-D_{13}+D_{22}+D_{23}+D_{24}-D_{25}
     -2 D_{26}-D_{310}+D_{36}\\
&&     -D_{38}+D_{39})[p_{2}, -p_{4}, -p_{3}, m_{b}, M_{s}, M_{s}, M_{s}]
-12 m_{t} (D_{12}-D_{13}+D_{22}\\
&&+D_{24}-D_{25}-D_{26}
-D_{310}+D_{36})[-p_{1}, p_{3}, -p_{2}, m_{b}, M_{s}, M_{s}, m_{b}]\\
&& +4 m_{t} (D_{12}-D_{13}+2 D_{24}
-2 D_{26}-D_{310}+D_{34})[p_{3}, -p_{1}, -p_{2}, m_{b}, m_{b}, M_{s}, m_{b}],
\end{array}
\eqno{(A.25)}
$$

$$
\begin{array}{lll}
f_{12}& = &f'_{12}+
32 m_{t}( D_{22} + D_{23} - D_{25} - D_{26} - D_{310} + D_{36}\\
&&- D_{38} + D_{39}
         )[p_{1}, -p_{3}, -p_{4}, M_{s}, m_{t}, m_{t}, m_{t}],
\end{array}
\eqno{(A.26)}
$$

$$
\begin{array}{lll}
f'_{13}& = &
 36 m_{t} (D_{310}-D_{37}-D_{38}+D_{39})[p_{2}, -p_{4}, -p_{3}, m_{b}, M_{s}, M_{s}, M_{s}]\\
&&+12 m_{t} (D_{310}-D_{38})[-p_{1}, p_{3}, -p_{2}, m_{b}, M_{s}, M_{s}, m_{b}]\\
&& -4 m_{t} (D_{310}-D_{35})[p_{3}, -p_{1}, -p_{2}, m_{b}, m_{b}, M_{s}, m_{b}],
\end{array}
\eqno{(A.27)}
$$

$$
\begin{array}{lll}
f_{13}& = &f'_{13}+
32 m_{t}( D_{25} - D_{26} + D_{310} - D_{37} - D_{38} \\
&&+ D_{39} )[p_{1}, -p_{3}, -p_{4}, M_{s}, m_{t}, m_{t}, m_{t}],
\end{array}
\eqno{(A.28)}
$$

$$
\begin{array}{lll}
f'_{14}& =&
 36 m_{t} (D_{22}-D_{24}
 +D_{25}-D_{26}-D_{34} \\
&&   +D_{35}+D_{36}-D_{37}-D_{38}+D_{39})[p_{2}, -p_{4}, -p_{3}, m_{b}, M_{s}, M_{s}, M_{s}]\\
&&+12 m_{t} (D_{23}-D_{26}-D_{38}+D_{39})[-p_{1}, p_{3}, -p_{2}, m_{b}, M_{s}, M_{s}, m_{b}]\\
&& -4 m_{t} (D_{23}-D_{25}-D_{35}+D_{37})[p_{3}, -p_{1}, -p_{2}, m_{b}, m_{b}, M_{s}, m_{b}],
\end{array}
\eqno{(A.29)}
$$

$$
f_{14} = f'_{14}+
32 m_{t}( D_{23} - D_{26} - D_{38} + D_{39}
         )[p_{1}, -p_{3}, -p_{4}, M_{s}, m_{t}, m_{t}, m_{t}],
\eqno{(A.30)}
$$

$$
\begin{array}{lll}
f'_{15}& =&
 12 a_{2} (
    m_{b}^2 C_{0}
    + m_{t}^2 (C_{11}+C_{21}-C_{12}-C_{23})
  ) [-p_{1}, p_{3}, m_{b}, M_{s}, M_{s}]\\
&&- 12 a_{2} (
  m_{b}^2 C_{0}-m_{t} m_{c} C_{12}-m_{t}^2 (C_{23}-C_{22})
   )[p_{3}, -p_{1}, m_{b}, m_{b}, M_{s}]\\
&&+6 (
m_{b}^2 D_{0}
+m_{t} m_{c} (D_{13}+D_{25})\\
&&+m_{t}^2 (D_{11}-D_{12}+D_{21}-D_{24})
   )[-p_{1}, p_{3}, -p_{2}, m_{b}, M_{s}, M_{s}, m_{b}]\\
&&- 2 (
m_{b}^2 D_{0} -m_{t} m_{c} (D_{12}-D_{13}-D_{26})\\
&&+m_{t}^2 (D_{22}+D_{24})
   )[p_{3}, -p_{1}, -p_{2}, m_{b}, m_{b}, M_{s}, m_{b}],
\end{array}
\eqno{(A.31)}
$$

$$
\begin{array}{lll}
f_{15}& =&
f'_{15}(m_{i}^2\rightarrow -m_{i}^2,~i=b,t)\\
&&+16 a_{2}m_{t}^2  (
 C_{11} + C_{21}-C_{12} - C_{23})
         )[p_{1}, -p_{3}, M_{s}, m_{t}, m_{t}]
+ 16m_{t} ( m_{c}   D_{13} \\
&&- m_{t} (D_{11} - D_{12} + D_{21} - D_{24} - D_{25} + D_{26})
         )[p_{1}, -p_{3}, -p_{4}, M_{s}, m_{t}, m_{t}, m_{t}],
\end{array}
\eqno{(A.32)}
$$

$$
\begin{array}{lll}
f'_{16}& = &
-2 (
m_{b}^2 (D_{11}-D_{12})+m_{t} m_{c} (D_{25}-D_{26})\\
&&-m_{t}^2 (D_{22}-D_{24})
   )[p_{3}, -p_{1}, -p_{2}, m_{b}, m_{b}, M_{s}, m_{b}],
\end{array}
\eqno{(A.33)}
$$

$$
\begin{array}{lll}
f_{16}& = &
f'_{16}(m_{i}^2\rightarrow -m_{i}^2, i=b,t)\\
&&+16m_{t}^2 (D_{11} - D_{12} + D_{21} - D_{24}-D_{25}
+ D_{26}) [p_{1}, -p_{3}, -p_{4}, M_{s}, m_{t}, m_{t}, m_{t}],
\end{array}
\eqno{(A.34)}
$$

$$
\begin{array}{lll}
f'_{17}& =&
6 (
m_{b}^2 D_{13} +m_{t} m_{c} D_{23}+m_{t}^2 (D_{25}-D_{26})
   )[-p_{1}, p_{3}, -p_{2}, m_{b}, M_{s}, M_{s}, m_{b}]\\
&& -2 (
m_{b}^2 D_{13}+m_{t} m_{c} D_{23}-m_{t}^2 (D_{25}-D_{26})
   )[p_{3}, -p_{1}, -p_{2}, m_{b}, m_{b}, M_{s}, m_{b}] ,
\end{array}
\eqno{(A.35)}
$$

$$
\begin{array}{lll}
f_{17}& =&
f'_{17}(m_{i}^2\rightarrow -m_{i}^2, i=b,t)\\
&&+16 m_{t}( m_{c} D_{13} + m_{t} (D_{25} - D_{26})
         )[p_{1}, -p_{3}, -p_{4}, M_{s}, m_{t}, m_{t}, m_{t}],
\end{array}
\eqno{(A.36)}
$$

$$
\begin{array}{lll}
f'_{18}& = &
 12 a_{1} (
   m_{b}^2 C_{0} - m_{t} m_{c} (C_{11}+C_{21})
   + m_{t}^2 (C_{12}+C_{23}) )[p_{2}, -p_{4}, m_{b}, M_{s}, M_{s}]\\
&&- 12 a_{1} (
    m_{b}^2 C_{0} -m_{t} m_{c} C_{22}
    +m_{t}^2 (C_{12}+C_{23}) )[-p_{4}, p_{2}, m_{b}, m_{b}, M_{s}]\\
&&-2 (
m_{b}^2 (D_{0}+D_{11})-m_{t} m_{c} (D_{23}-D_{25})\\
&&+m_{t}^2 (D_{12}-D_{13}+D_{24}-D_{26})
   )[p_{3}, -p_{1}, -p_{2}, m_{b}, m_{b}, M_{s}, m_{b}],
\end{array}
\eqno{(A.37)}
$$

$$
\begin{array}{lll}
f_{18}& = &
f'_{18}(m_{i}^2\rightarrow -m_{i}^2, i=b,t)\\
&&-16 a_{1} m_{t} (m_{c} (  C_{12} -
C_{21} + C_{23}) - m_{t} C_{11})[-p_{2}, p_{4}, M_{s}, m_{t}, m_{t}]\\
&& -16m_{t}^2  (D_{12} + D_{24}-D_{13} - D_{26})
         [p_{1}, -p_{3}, -p_{4}, M_{s}, m_{t}, m_{t}, m_{t}],
\end{array}
\eqno{(A.38)}
$$

$$\begin{array}{lll}
f'_{19} &=& 4a_{2} a_{3} m_{t} (m_{b}^2 B_{0} + m_{t}^2 B_{1})[-p_{1}, m_{b}, M_{s}]
+4a_{1} a_{3} m_{t} (m_{b}^2 B_{0} - m_{t} m_{c} B_{1})[p_{2}, m_{b}, M_{s}]\\
&&- 2a_{1} a_{2} m_{t} (
   m_{b}^2 B_{0} + m_{t}^2 B_{1} - 2 (p_{1}\cdot p_{3}) B_{1}
   )[-p_{1} + p_{3}, m_{b}, M_{s}]\\
&& -12 a_{2} m_{t} C_{24}[-p_{1}, p_{3}, m_{b}, M_{s}, M_{s}]
 -12 a_{1} m_{t} C_{24}[p_{2}, -p_{4}, m_{b}, M_{s}, M_{s}]\\
&&- 6 a_{2} m_{t} (2 C_{24}+m_{t}^2 (C_{12}-C_{22}) + 2 (p_{1}\cdot p_{3}) (C_{12}+C_{23}))
   [p_{3}, -p_{1}, m_{b}, m_{b}, M_{s}]\\
&&+ 6 a_{1} (
 2 m_{b}^2 m_{t} C_{0} - m_{t} (2 C_{24}-m_{c}^2 (C_{22}-C_{23})
 + m_{t} m_{c} C_{12} - m_{t}^2 (C_{11}+C_{21}))\\
&& + 2 m_{t} (p_{1}\cdot p_{2}-p_{2}\cdot p_{3}) (C_{11}-C_{12}+C_{21}-C_{23})\\
&&-2 m_{t} (p_{1}\cdot p_{3})  (C_{11}+C_{21})
   )[-p_{4}, p_{2}, m_{b}, m_{b}, M_{s}]\\
&& -18 m_{t} (D_{313}-D_{312}) D_{0}[p_{2}, -p_{4}, -p_{3}, m_{b}, M_{s}, M_{s}, M_{s}]
 -( m_{t} (4 D_{27}+4 D_{311}\\
&& +m_{b}^2 (D_{0}+D_{11})
 -m_{c}^2 (2  D_{23}+d37) -m_{t} m_{c} D_{13}
       -m_{t}^2 (D_{12}+2 D_{22}+d36)) \\
&&       +m_{b}^2 m_{t} D_{0}
-2  m_{t}  (p_{1}\cdot p_{2}  (D_{13}+2  D_{26}+D_{310})
              -p_{1}\cdot p_{3} (D_{12}+2  D_{24}+D_{34})\\
&&              -p_{2}\cdot p_{3} (D_{13}+2  D_{25}+D_{35}))
)[p_{3}, -p_{1}, -p_{2}, m_{b}, m_{b}, M_{s}, m_{b}]\\
&& -6 m_{t} (D_{27}-D_{312})[-p_{1}, p_{3}, -p_{2}, m_{b}, M_{s}, M_{s}, m_{b}]
 \end{array}
\eqno{(A.39)}
 $$

$$
\begin{array}{lll}
f_{19}& =&
f'_{19}(m_{t}m_{c}\rightarrow -m_{t}m_{c})
-8 a_{2} a_{3} m_{t}^3  B_{1}[-p_{1}, m_{t}, M_{s}]
+  8 a_{1} a_{2} m_{t}(2 p_{1} \cdot p_{3}\\
&&      - m_{t}^2) B_{1}[-p_{1} + p_{3}, m_{t}, M_{s}]
+ 8 a_{1} a_{3} m_{t}^2 m_{c} B_{1}[p_{2}, m_{t}, M_{s}]\\
&&-8 a_{2} m_{t} ( 2 C_{24} + m_{t}^2 ( C_{0} -
        C_{11} - C_{21}) \\
&&  + 2 (p_{1} \cdot p_{3}) (C_{12} + C_{23}) )[p_{1}, -p_{3}, M_{s}, m_{t}, m_{t}]
- 8 a_{1}m_{t}  ( 2 C_{24} - m_{c}^2 ( C_{21}  - 2 C_{23}) \\
&&        + m_{c} m_{t} C_{11} - m_{t}^2 (C_{12} + C_{22} - C_{0})
        - 2 (p_{1} \cdot p_{2}-p_{2} \cdot p_{3}) (C_{22} - C_{23})\\
&&        + 2 (p_{1} \cdot p_{3}) (C_{12} + C_{22})
         )[-p_{2}, p_{4}, M_{s}, m_{t}, m_{t}]
- 8 m_{t}(4 D_{312} - 4 D_{313} \\
&&+ m_{c}^2  D_{33}
 - m_{t}^2 ( D_{0} -
         D_{11} + D_{13} - D_{21} + D_{23} + 2 D_{24} - 2 D_{26} - 2 D_{310}\\
&&         - D_{33} + D_{34} - D_{35} + 2 D_{37}  + D_{39}) -
         2 (p_{1} \cdot p_{2}) (D_{25} - D_{26} - D_{310} - D_{33} +
         D_{37}  + D_{39})\\
&&     + 2 (p_{1} \cdot p_{3}) (D_{22} + D_{23} - 2 D_{26} - 2 D_{310} -
         D_{33} + D_{36} + D_{37} - D_{38} + 2 D_{39}) \\
&&         - 2 (p_{2} \cdot p_{3}) (D_{33} + D_{38} - 2 D_{39})
         )[p_{1}, -p_{3}, -p_{4}, M_{s}, m_{t}, m_{t}, m_{t}],
\end{array}
\eqno{(A.40)}
$$
$$
\begin{array}{lll}
f'_{20} & =&
 -18 m_{t} (D_{313}-D_{312})[p_{2}, -p_{4}, -p_{3}, m_{b}, M_{s}, M_{s}, M_{s}]\\
&&-6 m_{t} D_{312}[-p_{1}, p_{3}, -p_{2}, m_{b}, M_{s}, M_{s}, m_{b}]
+ 2 m_{t} D_{311}[p_{3}, -p_{1}, -p_{2}, m_{b}, m_{b}, M_{s}, m_{b}],
\end{array}
\eqno{(A.41)}
$$

$$
f_{20}  =f'_{20}+
16 m_{t}( D_{27} + D_{312} - D_{313}
         )[p_{1}, -p_{3}, -p_{4}, M_{s}, m_{t}, m_{t}, m_{t}],
\eqno{(A.42)}
$$

where
$$a_{1}=\frac{1}{\hat{t}-m_{t}^{2}},~~
a_{2}=\frac{1}{\hat{t}-m_{c}^{2}}
~~and ~~a_{3}=\frac{1}{m_{t}^{2}-m_{c}^{2}}.$$

$$ M^{\hat{u}}=M^{\hat{t}} ~~~(p_{3}\leftrightarrow p_{4}, ~~\mu \leftrightarrow
\nu, \hat{t}\leftrightarrow \hat{u})~~~~~~~(A.43)$$

\vskip 10mm
\begin{flushleft} {\bf Figure Captions} \end{flushleft}

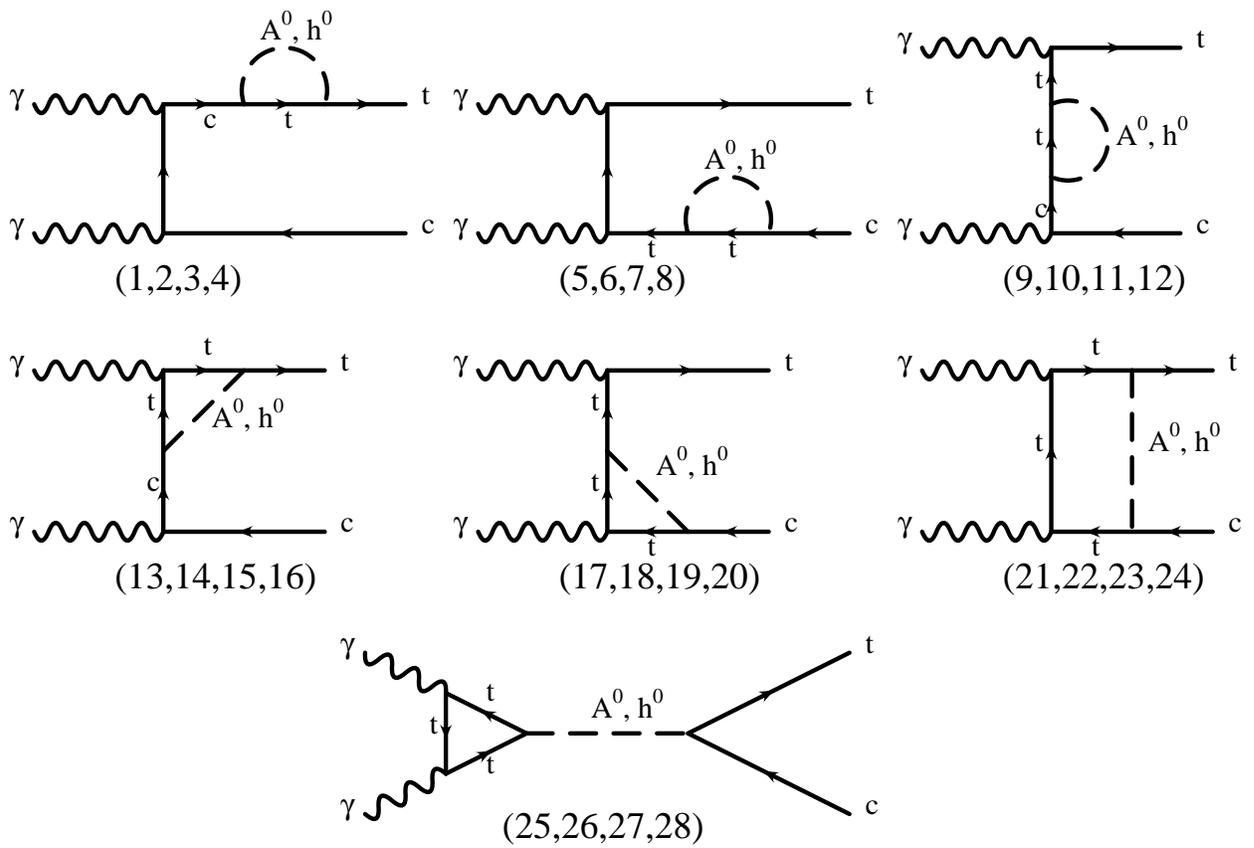
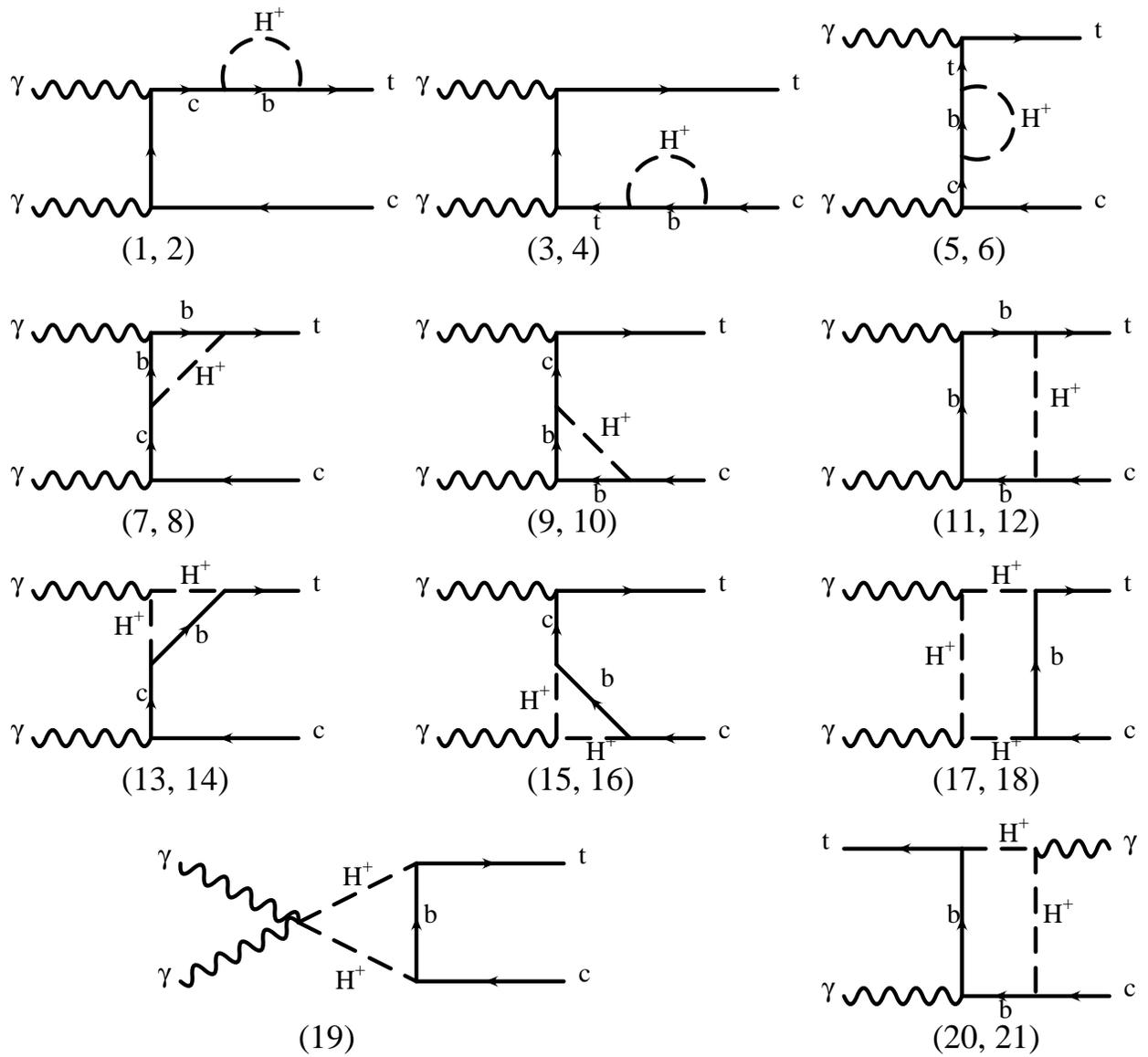
{\bf Fig.1} The Feynman diagrams of the subprocess $\gamma\gamma \rightarrow
t\bar{c}$.

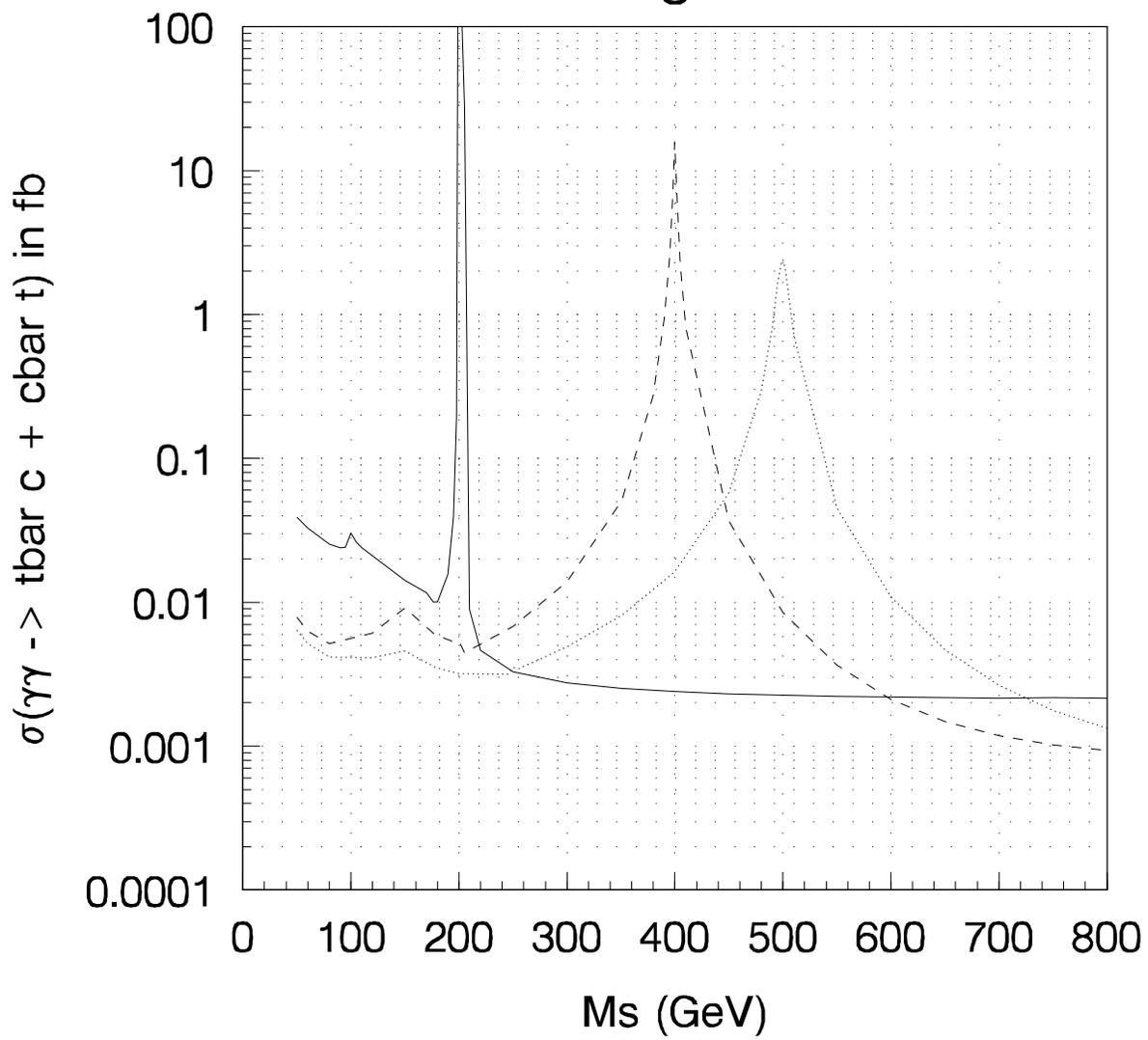
{\bf Fig.2} Total cross sections of the subprocess $\gamma\gamma \rightarrow
t\bar{c} + \bar{t}c$ as function of $M_{s}$. The solid curve is for
$\sqrt{\hat{s}}=200 GeV$, the dashed curve is for
$\sqrt{\hat{s}}=400 GeV$ and the dotted curve is for
$\sqrt{\hat{s}}=500 GeV$.

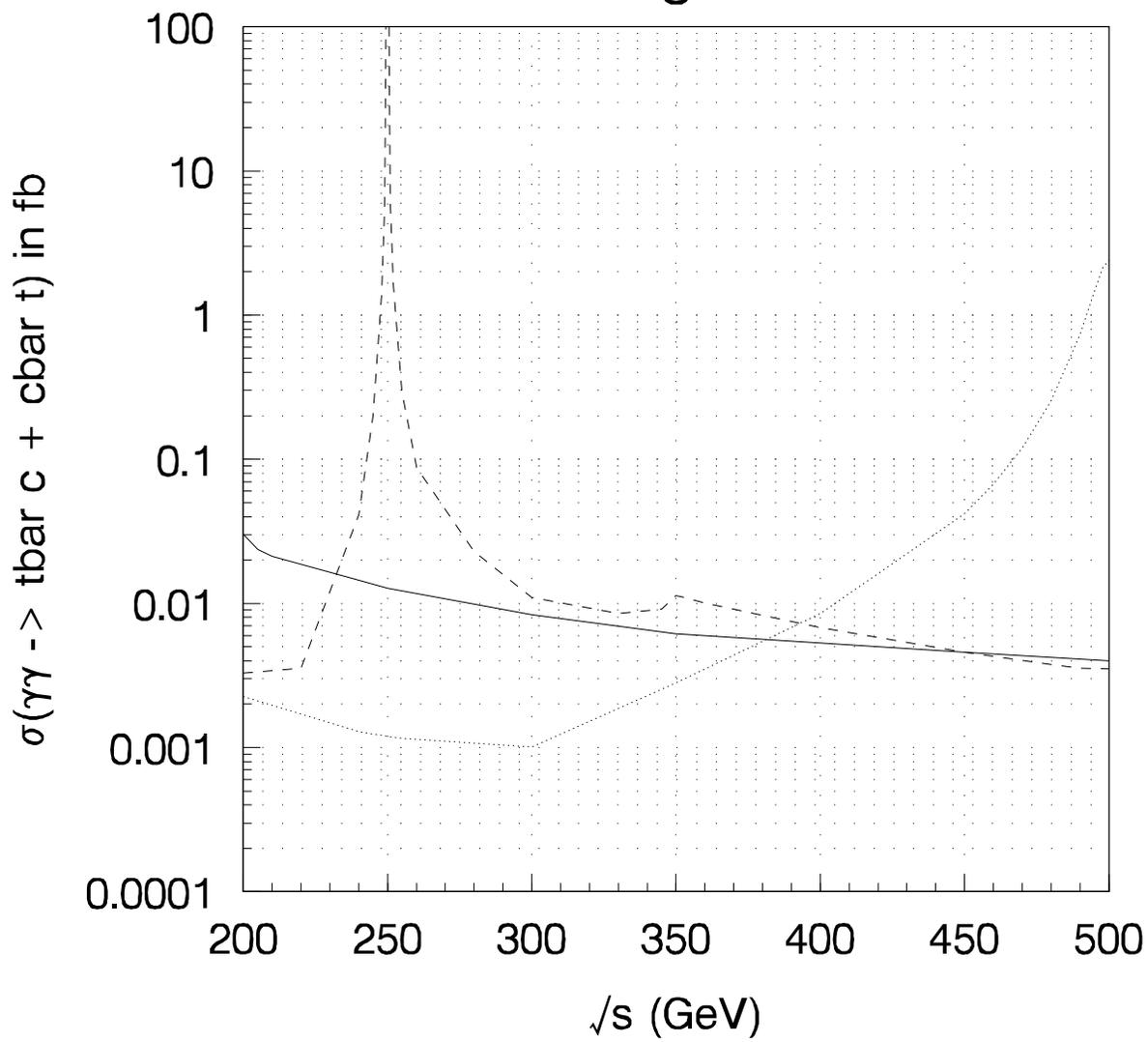
{\bf Fig.3} Total cross sections of the subprocess $\gamma\gamma \rightarrow
t\bar{c} + \bar{t}c$ as function of $\sqrt{\hat{s}}$. The solid curve is for
$M_{s}=100~GeV$, the dashed curve is for $M_{s}=250~GeV$ and the
dotted curve is for $M_{s}=500~GeV$.

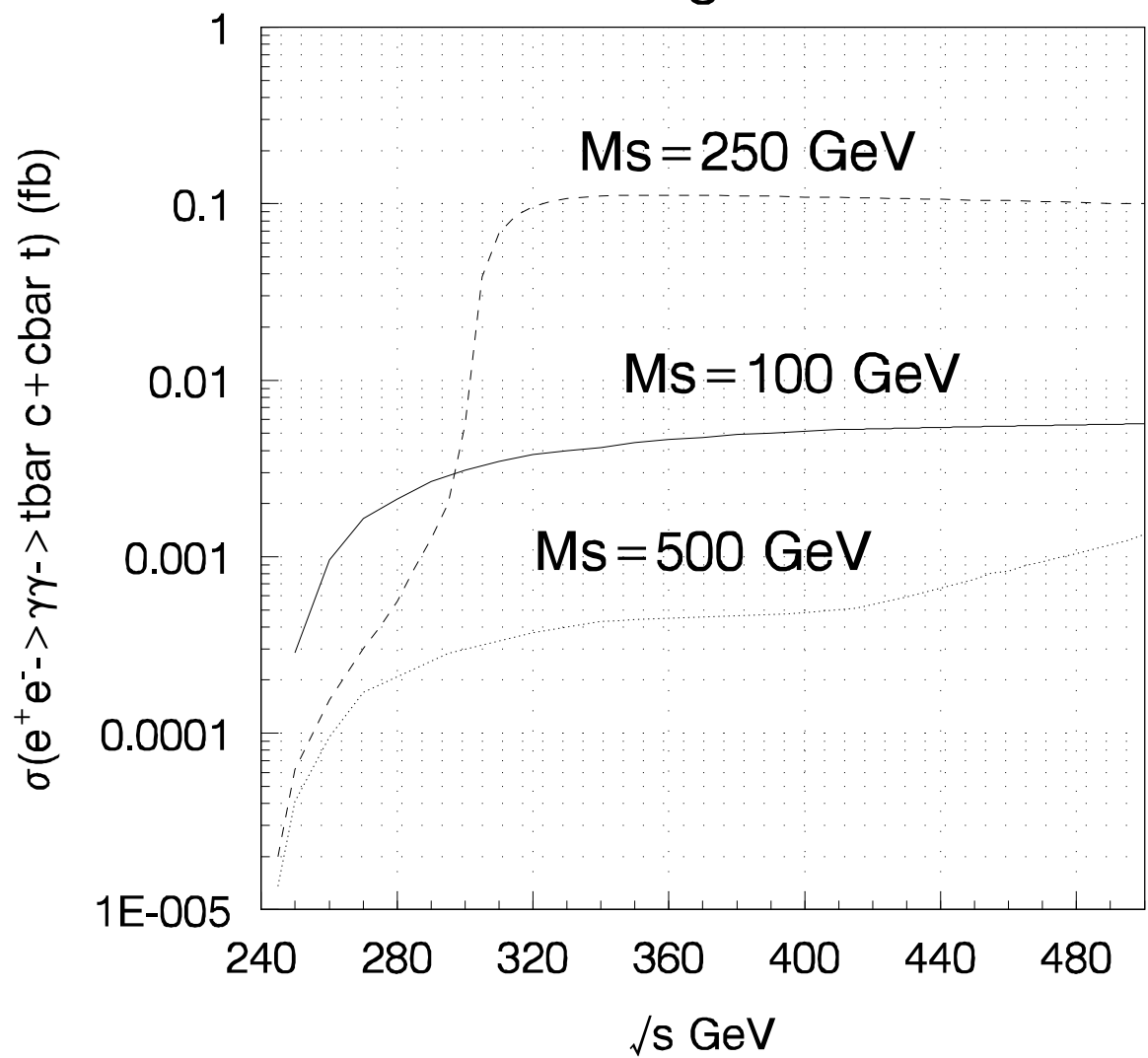
{\bf Fig.4} Total cross sections of the process $e^{+}e^{-}\rightarrow
\gamma\gamma \rightarrow t\bar{c} + \bar{t}c$ as function of $\sqrt{s}$.
The solid curve is for $M_{s}=100~GeV$, the dashed curve is for
$M_{s}=250~GeV$ and the dotted curve is for $M_{s}=500~GeV$.

\end{large}
\end{document}